\documentclass[preprint,prb]{revtex4}

\usepackage{graphicx}
\usepackage{dcolumn}
\usepackage{float}
\usepackage{amsmath}

\begin{document}

\title{\boldmath Temperature-dependent $f$-electron evolution in CeCoIn$_5$ via a comparative infrared study with LaCoIn$_5$ \unboldmath}

\author{Myounghoon Lee$^{1}$} \author{Yu-Seong Seo$^{1}$} \author{Seulki Roh$^{1}$} \author{Seokbae Lee$^{1}$} \author{Jihyun Kim$^{1}$} \author{Junwon Kim$^{2}$} \author{Tuson Park$^{1}$} \author{Ji Hoon Shim$^{3}$} \author{Jungseek Hwang$^{1}$}\email{jungseek@skku.edu}

\affiliation{$^{1}$Department of Physics, Sungkyunkwan University, Suwon, Gyeonggi-do 16419, Republic of Korea \\ $^2$Division of Advanced Materials Science, Pohang University of Science and Technology, Pohang, Gyeongbuk-do, 37673, Republic of Korea \\ $^3$Division of Advanced Materials Science, Pohang University of Science and Technology, Pohang, Gyeongbuk-do, 37673, Republic of Korea}


\begin{abstract}

We investigated CeCoIn$_5$ and LaCoIn$_5$ single crystals, which have the same HoCoGa$_5$-type tetragonal crystal structure, using infrared spectroscopy. However, while CeCoIn$_5$ has 4$f$ electrons, LaCoIn$_5$ does not. By comparing these two material systems, we extracted the temperature-dependent electronic evolution of the $f$ electrons of CeCoIn$_5$. We observed that the differences caused by the $f$ electrons are more obvious in low-energy optical spectra at low temperatures. We introduced a complex optical resistivity and obtained a magnetic optical resistivity from the difference in the optical resistivity spectra of the two material systems. From the temperature-dependent average magnetic resistivity, we found that the onset temperature of the Kondo effect is much higher than the known onset temperature of Kondo scattering ($\simeq$ 200 K) of CeCoIn$_5$. Based on momentum-dependent hybridization, the periodic Anderson model, and a maximum entropy approach, we obtained the hybridization gap distribution function of CeCoIn$_5$ and found that the resulting gap distribution function of CeCoIn$_5$ was mainly composed of two (small and large) components (or gaps). We assigned the small and large gaps to the in-plane and out-of-plane hybridization gaps, respectively. We expect that our results will provide useful information for understanding the temperature-dependent electronic evolution of $f$-electron systems near Fermi level.
\\


\noindent *Correspondence author \\
  Email: jungseek@skku.edu (Jungseek Hwang)

\end{abstract}

\maketitle

\section*{Introduction}

Heavy-fermion materials, which are quantum materials with intriguing properties, have been intensively investigated since their discovery \cite{steglich:1979,bonn:1988,donovan:1997,degiorgi:1999,petrovic:2001,dordevic:2001,dressel:2002,hancock:2004,silhanek:2006,park:2008,nagel:2012,wirth:2016,chen:2016,kirchner:2020}. Heavy-fermion materials contain $f$ electrons and exhibit interesting temperature-, doping-, and pressure-induced phase transitions, attributable to two major competing interactions. The Doniach phase diagram provides a qualitative interpretation of the possible ground states in the heavy-fermion systems \cite{doniach:1977}, which can be stabilized by two competing interactions, namely, the Kondo and the Ruderman, Kittel, Kasuya, and Yosida (RKKY) interactions. The two interactions can be controlled by a tuning parameter: $JD(E_F)$, where $J$ is the exchange coupling constant between localized and itinerant spins, $D(E_F)$ is the density of states at the Fermi level, and $E_F$ is the Fermi energy. The Kondo interaction can be described in terms of the parameter as $T_K \propto D(E_F)^{-1} \exp{[-1/JD(E_F)]}$, where $T_K$ is the single impurity Kondo temperature, which is the crossover temperature between a logarithmic Kondo scattering regime and a nonperturbative strong Kondo coupling regime \cite{jang:2020}. Similarly, the RKKY interaction can be described as $T_{RKKY} \propto J^2D(E_F)$, which represents interacting two localized spins by exchanging itinerant spins, where $T_{RKKY}$ is the energy scale of the RKKY interaction \cite{yang:2008}. When the energies of the two interactions are the same, that is, at a critical tuning parameter ($[JD]_c$), a quantum critical point may occur. Below (above) the critical tuning parameter, the RKKY (Kondo) interaction is dominant. Some of the possible external tuning parameters are pressure, magnetic field, and doping \cite{knebel:2001,silhanek:2006,park:2008}. CeCoIn$_5$ is known to be located near the quantum critical point in the phase diagram of the heavy-fermion 115 series (Ce$M$In$_5$ with $M$ = Co, Ir, or Rh) \cite{sidorov:2002}. CeCoIn$_5$ exhibits interesting temperature-dependent DC resistivity and optical properties \cite{petrovic:2001,singley:2002,shishido:2002,mena:2005,shim:2007,burch:2007,jang:2020}. There were a couple of optical studies of CeCoIn$_5$ comparing with CeRhIn$_5$ and CeIrIn$_5$ have been reported \cite{mena:2005,burch:2007}. There are two characteristic temperatures (or energy scales) associated with the DC resistivity of the CeCoIn$_5$ system \cite{yang:2008,jang:2020}: the single-ion Kondo temperature ($T_K$) and the Kondo lattice coherent temperature ($T^*$), which is closely related to the intersite RKKY interaction ($T_{RKKY}$) of a Kondo lattice material \cite{yang:2008}. A study showed that below $T^*$, the Kondo gas is partially condensed into a heavy electron Kondo liquid that has a temperature-independent Wilson ratio = 2.0\cite{nakatsuji:2003}. The Kondo scattering caused by the $f$ electrons begins at a higher temperature than the two aforementioned characteristic temperatures ($T_K$ and $T^*$), the onset temperature of which is denoted as $T_K'$. The $T_K'$ can be estimated from the resistivity minimum, where $-\ln(T)$-dependent DC resistivity appears \cite{jang:2020}. Chen {\it et al.} studied CeCoIn$_5$ using high-resolution angle-resolved photoemission spectroscopy (ARPES) and revealed that the localized-to-itinerant transition takes place at very high temperatures, where $f$ electrons are still largely localized \cite{chen:2017}. Another recent study on CeCoIn$_5$ that combined dynamic mean-field theory (DMFT) and ARPES revealed that the onset temperature of the $f$-electron effect is higher than the known onset temperature of Kondo scattering, $T_K'$ \cite{jang:2020}. The intriguing temperature-dependent electronic structure evolution of CeCoIn$_5$ has been an important research topic since its discovery \cite{petrovic:2001}. Infrared and optical spectroscopy have long been recognized as a quantitative and practical experimental technique for studying electron and phononic transitions in target material systems.

In this study, we investigated single crystals of CeCoIn$_5$ and LaCoIn$_5$ using infrared spectroscopy. By comparing the optical spectra of these two material systems, we obtained temperature-dependent electronic properties of the $f$ electrons. We note that there has been a similar comparative study of CeCoIn$_5$ and LaCoIn$_5$ that uses a photoemission technique \cite{chen:2019}. We observe that CeCoIn$_5$ exhibits non-monotonic temperature-dependent changes in the low-frequency region, as in the reported optical spectra of CeCoIn$_5$ \cite{singley:2002,mena:2005}, while LaCoIn$_5$ undergoes a monotonic temperature-dependent change in the entire measured spectral range, similar to that observed in a typical metal. At low temperatures \cite{shim:2007}, the two material systems gave rise to significantly different spectra. However, at room temperature, the spectra of both materials were quite similar. To compare these two material systems, we introduce the complex optical resistivity, which is defined by the inverse of the complex optical conductivity. We also introduced another optical quantity, magnetic optical resistivity, which is defined by the difference between the optical resistivity spectra of the two materials. Since the magnetic optical resistivity is additional resistivity caused by the $f$ electrons, it is intimately associated with the Kondo (or $f$ electron) effect. As the temperature rises, the magnetic optical resistivity decreases. We found that the onset temperature of the average magnetic resistivity is much higher than the known onset temperature of Kondo scattering ($T_K' \cong 200 K$) of CeCoIn$_5$. Our results are consistent with an ARPES study by Chen {\it et al.} \cite{chen:2017} and a recent combined study of DMFT and ARPES \cite{jang:2020}. We also obtained the hybridization gap distribution function of CeCoIn$_5$ using the proposed approach by Burch {\it et al.} \cite{burch:2007} through a maximum entropy method. The obtained gap distribution function mainly consisted of two (small and large) gaps. Based on the previous LDA+DMFT study \cite{shim:2007}, we assigned the small and large gaps to the in-plane and out-of-plane hybridization gaps, respectively. We also used an extended Drude model \cite{gotze:1972,allen:1977,puchkov:1996,hwang:2004} to obtain the optical effective mass. As we expected, the two material systems exhibited significantly different temperature-dependent effective masses.

\section*{Experiments}

\begin{figure}[!htbp]
  \vspace*{-0.8 cm}%
  \centerline{\includegraphics[width=6 in]{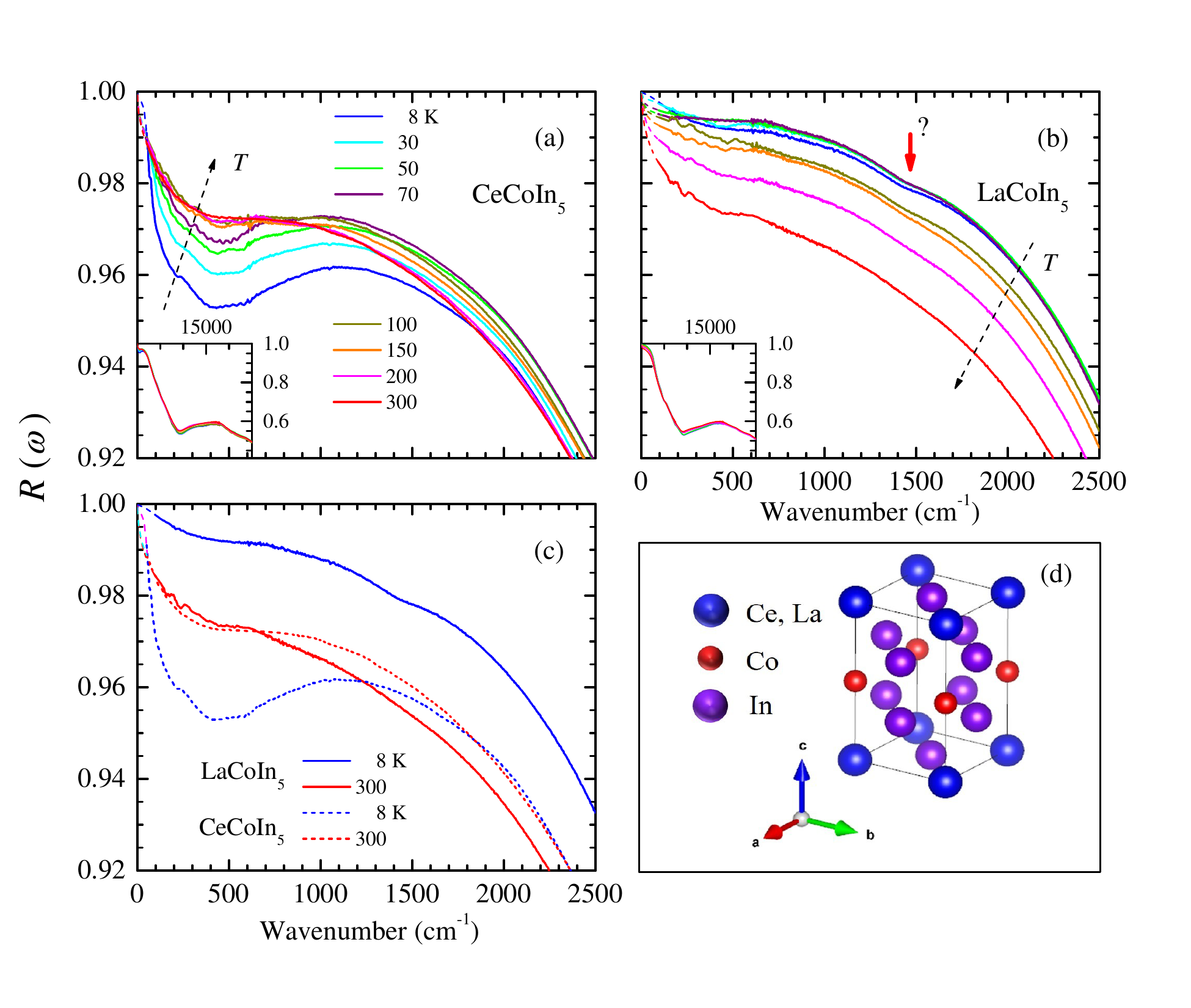}}%
  \vspace*{-0.8 cm}%
\caption{Measured reflectance spectra of (a) CeCoIn$_5$ and (b) LaCoIn$_5$ at various temperatures. The low frequency extrapolations are shown as dashed lines. In the insets, the same spectra are shown in a wider spectral range of up to 25000 cm$^{-1}$. (c) Reflectance spectra of CeCoIn$_5$ and LaCoIn$_5$ at 8 and 300 K for comparison. (d) Common crystal structure of CeCoIn$_5$ and LaCoIn$_5$.}
 \label{fig1}
\end{figure}

High-quality single crystals of CeCoIn$_5$ and LaCoIn$_5$, which have a HoCoGa$_5$ type tetragonal structure (see Fig. \ref{fig1}(d)), were grown using the In self-flux method. The samples grown by the In self-flux method often contain indium. We removed the indium by using a centrifuge. The remnant indium was further removed by dissolving the indium with diluted hydrochloric acid \cite{hu:2013}. Then, we selected samples with smooth and flat surfaces. We did not polish the sample surface. The lattice constants of CeCoIn$_5$ (LaCoIn$_5$) are $a =$ 4.62 (4.64) \AA $\:$and $c =$ 7.56 (7.62) \AA. Details of the growth method can be found in the literature \cite{petrovic:2001,macaluso:2002}. The areas of the CeCoIn$_5$ and LaCoIn$_5$ samples used in this optical study were approximately 3$\times$3 and 1.5$\times$1.5 mm$^{2}$, respectively. We measured near-normal reflectance spectra of the $ab$ plane of CeCoIn$_5$ and LaCoIn$_5$ over a wide spectral range from 50 to 25,000 cm$^{-1}$ using a commercial Fourier transform infrared spectrometer (Vertex 80v, Bruker, Germany). To investigate the temperature-dependent properties, we took spectra at several controlled equilibrium temperatures from 8 to 300 K using a continuous liquid He flow cryostat. To obtain accurate reflectance spectra, we employed an {\it in-situ} metallization technique \cite{homes:1993}. Figs. \ref{fig1}(a) and (b) show the measured reflectance spectra of the two single-crystal samples at various temperatures. Temperature-dependent changes occurred mostly in the low-frequency region. The insets of Figs. \ref{fig1}(a) and (b) show the reflectance spectra of the two materials over a wide spectral range up to 25,000 cm$^{-1}$. The overall reflectance spectra of the two materials were similar. However, considerable differences between them were observed in the far- and mid-infrared regions at low temperatures. The existence of 4$f$ electrons in CeCoIn$_5$ may cause these differences. As the temperature decreases, the dip in the reflectance of CeCoIn$_5$ observed in the vicinity of 400 cm$^{-1}$ becomes deeper, which is consistent with other previous optical studies of CeCoIn$_5$ \cite{singley:2002,mena:2005}. On the other hand, as the temperature is lowered, the reflectance of LaCoIn$_5$ increases monotonically, which is the usual temperature-dependent behavior of typical metals. One interesting feature marked with a red arrow at near 1500 cm$^{-1}$ appears at low temperatures. Below 1500 cm$^{-1}$, the reflectance slightly is suppressed. But, we cannot find out the origin of this feature yet. In Fig. \ref{fig1}(c), we compare the reflectance spectra of the two materials in the low-frequency region below 2500 cm$^{-1}$ at two temperatures, 8 and 300 K. At 300 K, the two reflectance spectra are slightly different. However, at 8 K, the two reflectance spectra are considerably different, which indicates that the 4$f$ electrons in CeCoIn$_5$ are significantly temperature-dependent.

We employed a Kramers-Kronig analysis to obtain the optical conductivity from the measured reflectance spectrum. For performing the Kramers-Kronig analysis, the measured spectrum in a finite spectral range must be extrapolated to both zero and infinity frequencies. For the extrapolation to zero, we used the Hagen-Rubens relation, that is, $1-R(\omega) \propto \omega^{1/2}$. For the extrapolation to infinity, we assumed that $R(\omega) \propto \omega^{-2}$ from 25,000 to 10$^6$ cm$^{-1}$ and the free electron behavior, that is, $R(\omega) \propto \omega^{-4}$, above 10$^6$ cm$^{-1}$. We note that, in the high-frequency region above $\sim$10,000 cm$^{-1}$, we obtained a level of the optical conductivity for CeCoIn$_5$ comparable to that obtained in a previous ellipsometry study \cite{mena:2005}.

\section*{Results and discussions}

\subsection*{Optical conductivity}

\begin{figure}[!htbp]
  \vspace*{-0.8 cm}%
  \centerline{\includegraphics[width=6 in]{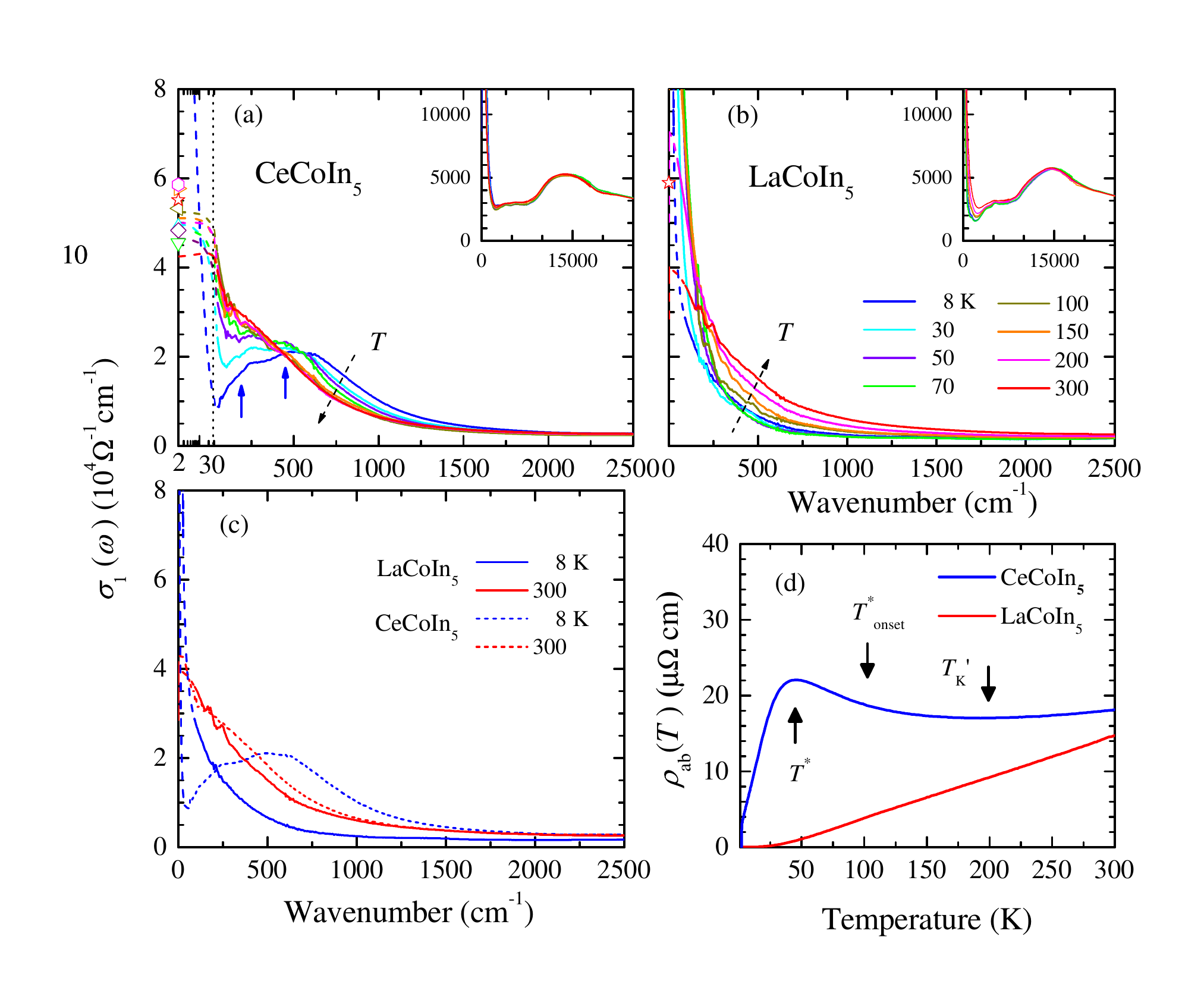}}%
  \vspace*{-0.8 cm}%
\caption{Optical conductivity of (a) CeCoIn$_5$ and (b) LaCoIn$_5$ at various temperatures. On the vertical axes, we marked the measured DC conductivity data points, which are the inverses of the measured DC resistivity data points, with open symbols. The low-frequency extrapolations are shown as dashed lines. In the insets, the same spectra in a wider spectral range up to 25000 cm$^{-1}$. (c) Conductivity spectra of CeCoIn$_5$ and LaCoIn$_5$ at 8 and 300 K for comparison. (d) Measured DC resistivity data of CeCoIn$_5$ and LaCoIn$_5$ as functions of temperature.}
 \label{fig2}
\end{figure}

\begin{figure}[!htbp]
  \vspace*{-0.3 cm}%
  \centerline{\includegraphics[width=6.5 in]{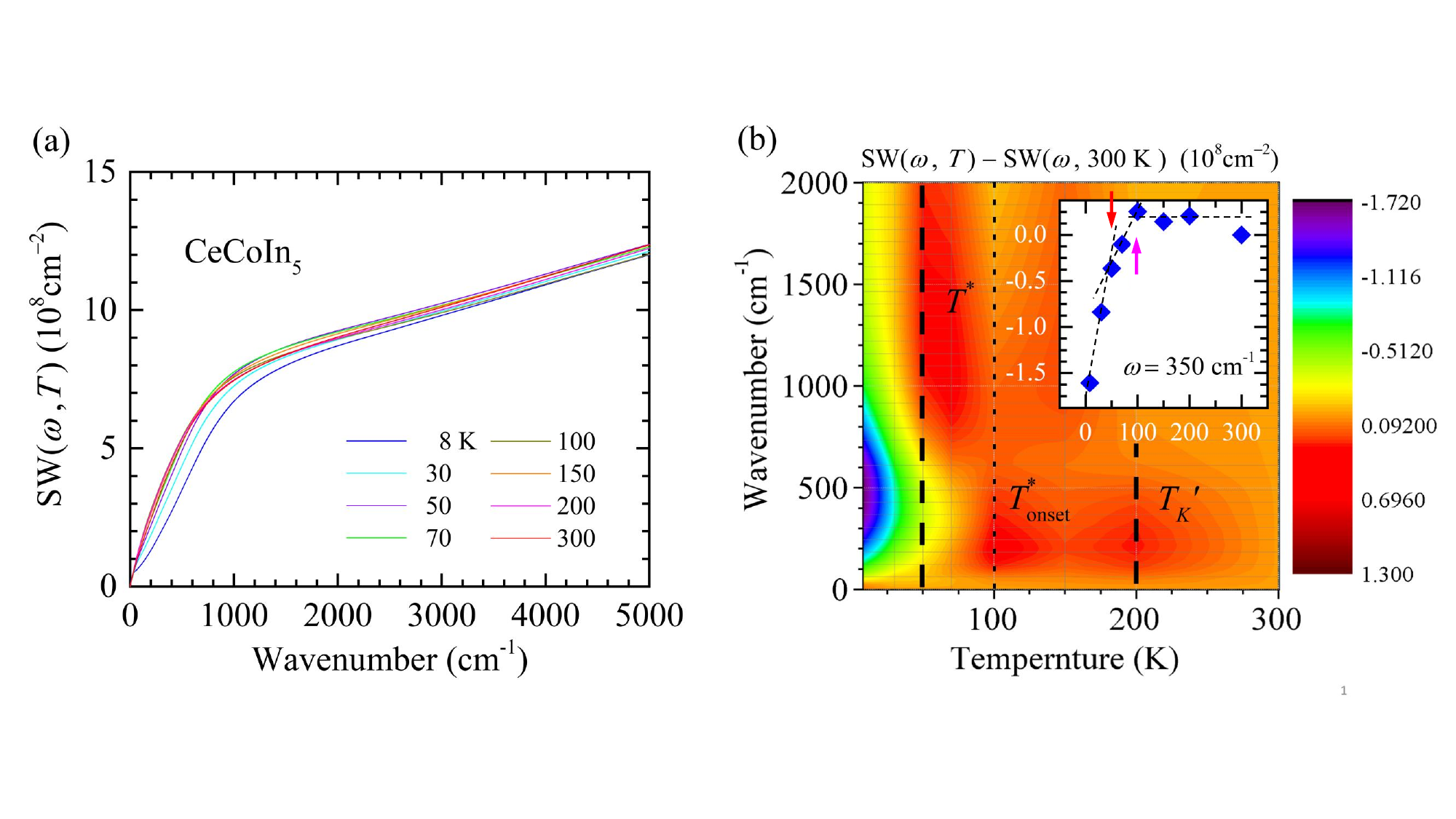}}%
  \vspace*{-1.5 cm}%
\caption{(a)Accumulated spectral weight ($SW(\omega)$) of CeCoIn$^5$ at various temperatures. (b) Color map of $\Delta SW(\omega, T)$. In the inset, $\Delta SW(\omega, T)$ is plotted at $\omega = $ 350 cm$^{-1}$. The red and magenta vertical arrows correspond to $T^*$ and $T^*_{\mathrm{onset}}$, respectively.}
 \label{fig3}
\end{figure}

Figs. \ref{fig2}(a) and (b) show the optical conductivity spectra of CeCoIn$_5$ and LaCoIn$_5$ at various temperatures, respectively. The optical conductivity of CeCoIn$_5$ was similar to previously reported ones \cite{singley:2002,mena:2005}. We clearly observed at least two hybridization gaps below $\sim$ 600 cm$^{-1}$ at the lowest temperature (8 K), which are marked with two vertical arrows (see Fig. \ref{fig5} and the related text for a more detailed discussion of the gaps). These gaps become more pronounced as the temperature decreases from 300 K. As lowering the temperature from 300 K, a dip near 150 cm$^{-1}$, which is intimately associated with the hybridization gap, starts to deepen near 100 K and the deepening rate significantly enhances near the lattice coherence temperature, $T^*$ ($\sim$50 K) (see more detailed descriptions with an accumulated spectral weight in Fig. \ref{fig3}). As we expected from the measured reflectance spectra and DC resistivity data (Fig. \ref{fig2}(d)), CeCoIn$_5$ exhibits a dramatic change with lowering the temperature, while LaCoIn$_5$ exhibits a monotonic temperature-dependent change. As we can see in the insets, the overall conductivity spectra of the two materials are quite similar, although they are not exactly the same. In Fig. \ref{fig2}(c), we compare the conductivity spectra of the two materials at two temperatures, 8 and 300 K. At 300 K, the conductivity spectra of the two samples show a small difference, as expected from the measured reflectance spectra. However, at 8 K, they are significantly different from each other; while LaCoIn$_{5}$ shows a narrower Drude-like mode compared with the optical conductivity at 300 K, CeCoIn$_5$ shows a much narrower Drude-like mode nearby zero frequency and gaps below 500 cm$^{-1}$.

In Fig. \ref{fig2}(d), we show the measured DC resistivity data of the two samples as functions of temperature. The resistivity of CeCoIn$_5$ exhibits several characteristic temperatures. As the temperature was lowered from room temperature (300 K), the DC resistivity initially decreases and shows a broad dip near 200 K. Below the dip, it gradually increases and then more rapidly increases near 100 K, resulting a slope change, and a rapid decrease occurs below 45 K, resulting in a distinct peak near 45 K. We note that the initial decrease results from the metallic behavior caused by the conduction electrons. The temperatures at the dip, the slope change, and peak are known as the onset temperatures of the Kondo scattering ($T_K'$), the onset temperature ($T^*_{\mathrm{onset}}$) of the Kondo lattice coherence, and the Kondo lattice coherence ($T^*$), respectively\cite{jang:2020}. Below $T_K'$, the Kondo scattering is dominant, where the conduction electrons are interrupted by the spins of the 4$f$ electrons, resulting in an increase in the resistivity versus temperature curve. The Kondo scattering becomes stronger as the temperature decreases. The coherence between screened 4$f$ electrons (or Kondo singlets) starts near $T^*_{\mathrm{onset}}$ and dominates below $T^*$, where the Kondo lattice sets in, and residual conduction electrons that are not involved in the Kondo screening effect is not influenced by the local spin lattice. They contribute almost freely to the conductivity with small scattering rates, resulting in a rapid decrease in the resistivity. However, a recent theoretical and experimental study using DMFT and ARPES for CeCoIn$_5$ suggested that the onset temperature of the Kondo effect is much higher than $T_K'$ ($\sim$ 200 K) \cite{jang:2020}. In fact, this is evident in the magnetic resistivity ($\rho_M$)\cite{nakatsuji:2002}, which is defined as $\rho_{\mathrm{M}}(T) \equiv \rho_{\mathrm{Ce}}(T) - \rho_{\mathrm{La}}(T)$, where $\rho_{\mathrm{Ce}}(T)$ and $\rho_{\mathrm{La}}(T)$ are the resistivities of CeCoIn$_5$ and LaCoIn$_5$, respectively (see the inset of Fig. \ref{fig7}(c)) because $\rho_M(T)$ has finite values above 200 K.

The accumulated spectral weight ($SW(\omega)$) can be defined by $SW(\omega,T)\equiv (120/\pi)\int_0^{\omega}\sigma_1(\omega',T)d\omega'$, where all frequencies are in cm$^{-1}$. The spectral weights of CeCoIn$_5$ at various temperatures are shown in Fig. \ref{fig3}(a). As the temperature decreases, the spectral weight transfers from a low-frequency region to a high-frequency one, resulting in the formation of the hybridization gap. To see the trend better, we take $\Delta SW(\omega, T) = SW(\omega, T) - SW(\omega, 300 K)$ and display them with a color map in Fig. \ref{fig3}(b). In the inset, we show $\Delta SW(\omega, T)$ at 350 cm$^{-1}$ as a function of temperature. As lowering the temperature, two characteristic temperatures are identified; one (magenta vertical arrow) is near 100 K, where $\Delta SW(\omega, T)$ starts to drop and the other (red vertical arrow) is near 50 K, where the dropping rate increases significantly, resulting in the appearance of a kink. The higher temperature ($\sim$100 K) is corresponding to the onset temperature of the lattice coherence ($T^*_{\mathrm{onset}}$) and the lower one ($\sim$50 K) corresponds to the lattice coherence temperature ($T^*$). We assigned the temperatures based on the characteristic temperatures in the DC resistivity.

\begin{figure}[!htbp]
  \vspace*{-0.3 cm}%
  \centerline{\includegraphics[width=6.0 in]{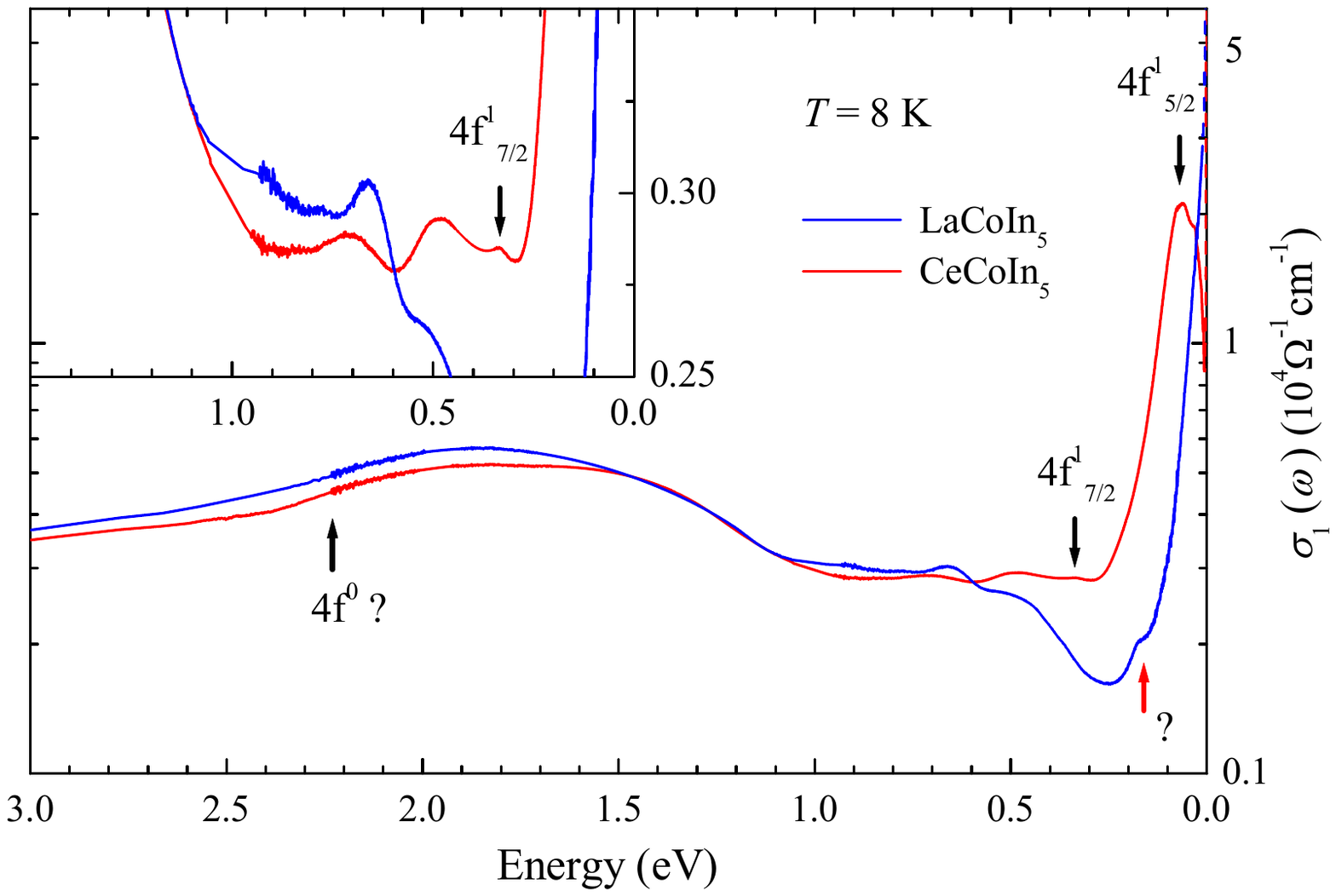}}%
  \vspace*{-1.0 cm}%
\caption{Optical conductivity of CeCoIn$_5$ and LaCoIn$_5$ in a wide spectral range at 8 K. In the inset, a magnified view of the optical conductivity spectra in a spectral range from 0.0 to 1.5 eV.}
 \label{fig4}
\end{figure}

We compared the optical conductivity spectra of CeCoIn$_5$ and LaCoIn$_5$ at 8 K in a wide spectral rage (0.0 - 3.0 eV or 0.0 - 24000 cm$^{-1}$) as shown in Fig. \ref{fig4}. Several sharp features are observed below 1.0 eV. We assigned a couple of them as 4$f^1_{5/2}$ and 4$f^1_{7/2}$ by comparing them with the angle-integrated EDC of CeCoIn$_5$ and LaCoIn$_5$ in Ref. [26]. However, the 4$f^0$ peak is not very clear in the optical spectrum. In the inset, we display a magnified view of the optical conductivity in a range between 0.0 and 1.5 eV to show those sharp and week features more clearly. The 4$f^1_{5/2}$ and 4$f^1_{7/2}$ sharp features clearly appear at low temperatures (see Fig. S1 in Supplementary Materials for more data at various temperatures). The 4$f^1_{5/2}$ is the same peak near 500 cm$^{-1}$ in Fig. \ref{fig2}(a), which is intimately associated with the hybridization gap.

\subsection*{Hybridization gaps}

\begin{figure}[!htbp]
  \vspace*{-0.3 cm}%
  \centerline{\includegraphics[width= 6.0 in]{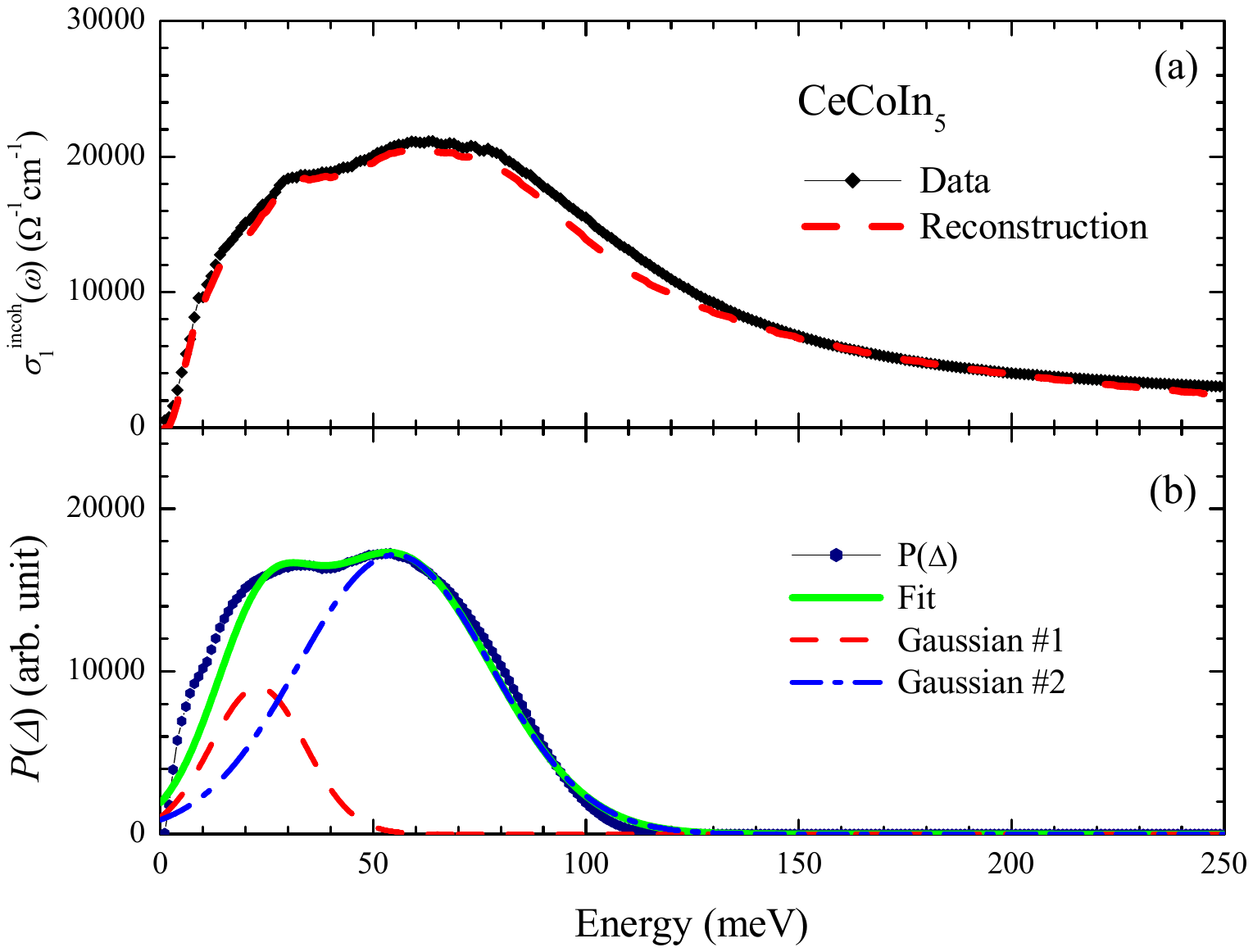}}%
  \vspace*{-0.7 cm}%
\caption{(a) $\sigma_1^{\mathrm{incoh}}(\omega)$ of CeCoIn$_5$ and reconstruction using the MEM at 8 K. (b) $P(\Delta)$ of CeCoIn$_5$ and a fit with two Gaussian modes.}
 \label{fig5}
\end{figure}

The strength of the hybridization between local moments (Ce 4$f$ electrons) and conduction electrons (In 5$p$ electrons) can be probed by optical spectroscopy. The strength appears as the hybridization gap size. The size of the gap may depend on the $k$-space point. But, optical measurement gives the $k$-averaged quantity. According to a previous LDA+DMFT study, we expect two dominant hybridization gaps \cite{shim:2007}. Two different-sized hybridization gaps are shown in Fig. \ref{fig2}(a) as marked with vertical arrows: one is around 190 cm$^{-1}$ and the other around 450 cm$^{-1}$ at 8 K. These two gaps have been identified as the in-plane (the smaller) and out-of-plane (the larger) gaps by Shim {\it et al.} using the LAD+DMFT calculations \cite{shim:2007}. The in-plane gap is caused by hybridization between Ce 4$f$ orbital and the in-plane In 5$p$ orbital, while the out-of-plane gap is caused by hybridization of Ce 4$f$ and the out-of-plane In 5$p$ orbital. We note that, in a previous optical study \cite{singley:2001}, the authors differently assigned the two features in the optical conductivity of CeCoIn$_5$: they assigned the feature at the lower energy as an intraband absorption coupled to a boson mode and the feature at the higher energy as the hybridization gap. However, the origin of the boson mode has not been found yet. We also note that the Ce atom exists as a Ce$^{3+}$ ion in CeCoIn$_5$. As a result, Ce's external 5$s$ and 5$d$ orbits cannot intervene in the hybridization. We estimated the hybridization gap distribution from measured optical conductivity using a similar approach introduced by Burch {\it et al.} \cite{burch:2007}, which is based on the periodic Anderson model and momentum-dependent hybridization. In the approach, the incoherent part of the optical conductivity ($\sigma_1^{incoh}(\omega)$), of which the Drude component (or the response of the coherent carriers) is removed, can be described as the following integral equation:
\begin{equation}\label{}
  \sigma_1^{incoh}(\omega)=\int_0^{\omega_c} P(\Delta)\sigma_1^{PAM}(\omega,\Delta) d\Delta
\end{equation}
where $\sigma_1^{PAM}(\omega,\Delta)$ is the optical conductivity of the periodic Anderson model \cite{dordevic:2001,hancock:2004} with a gap of $\Delta$, $P(\Delta)$ is the gap distribution function, and $\omega_c$ is the cutoff frequency. This is an inversion problem; we need to obtain $P(\Delta)$ from the measured $\sigma_1^{incoh}(\omega)$ with the known kernel, $\sigma_1^{PAM}(\omega,\Delta)= A\Theta(\omega-\Delta)/\sqrt{\omega^2-\Delta^2}$, where $A$ is a constant and $\Theta(x)$ is the Heaviside step function. To solve this inversion problem, we applied the maximum entropy method (MEM), which is a model-independent method and has been used for solving the inversion problems with the same mathematical form, i.e., generalized Allen’s formula \cite{allen:1971,schachinger:2006}. Fig. \ref{fig5}(a) shows $\sigma_1^{incoh}(\omega)$ of CeCoIn$_5$ and a reconstruction obtained using the MEM at 8 K. Fig. \ref{fig5}(b) shows the resulting $P(\Delta)$ of CeCoIn$_5$ and a fit with two Gaussian modes. From this study, we found that $P(\Delta)$ mainly consists of two Gaussian components. Each Gaussian mode may represent a corresponding hybridization gap with an energy dispersion. Based on the previous LDA+DMFT study \cite{shim:2007}, the Gaussian at the lower energy is assigned to the hybridization gap between the Ce 4$f$ orbital and the in-plane In 5$p$ orbital, while the one at the higher energy is assigned to the hybridization gap between the Ce 4$f$ orbital and the out-of-plane In 5$p$ orbital. In previous ARPES studies, three major ($\alpha$, $\beta$, and $\gamma$) bands were observed near the Fermi level in the $\Gamma$-M and X-M directions \cite{jang:2020,chen:2017}. The $\alpha$ and $\beta$ bands have approximately 2D Fermi surface character, consist mainly of the in-plane In 5$p$ orbital, and are partially influenced by the out-of-plane In orbital as well (see Fig. S2 in Supplementary Materials). These bands are gapped with a gap of $\sim$30 meV, as can be found in the previous studies \cite{jang:2020,chen:2017}. The $\gamma$ band, on the other hand, has a 3D Fermi surface character, is influenced by out-of-plane In orbitals, including Co-orbital (see Fig. S2 in Supplementary Materials), and has a gap of $\sim$70 meV, which exhibits a more complex temperature dependence. The peak position of each Gaussian is the average size of each hybridization gap. The average size of the in-plane (out-of-plane) hybridization gap is around 190 cm$^{-1}$ (450 cm$^{-1}$). Interestingly, the two Gaussian peaks exhibit different peak intensities. The 2D $\alpha$ and $\beta$ Fermi surfaces with the smaller in-plane gap have a much larger Fermi surface volume than the 3D $\gamma$ Fermi surface. However, their optical spectral weight is smaller than that of the $\gamma$ band. Therefore, the relationship between the Fermi surface volume and the optical weight seems to be non-trivial. We note that we have two Gaussian peaks for the gap distribution while Burch {\it et al.} had four Gaussian peaks in the distribution\cite{burch:2007}. We do not clearly understand what is different in these two analyses yet. We also note that our approach is model independent, whereas Burch {\it et al.}'s approach is model dependent. 

\subsection*{Optical resistivity and magnetic optical resistivity}

\begin{figure}[!htbp]
  \vspace*{-0.5 cm}%
  \centerline{\includegraphics[width= 6.0 in]{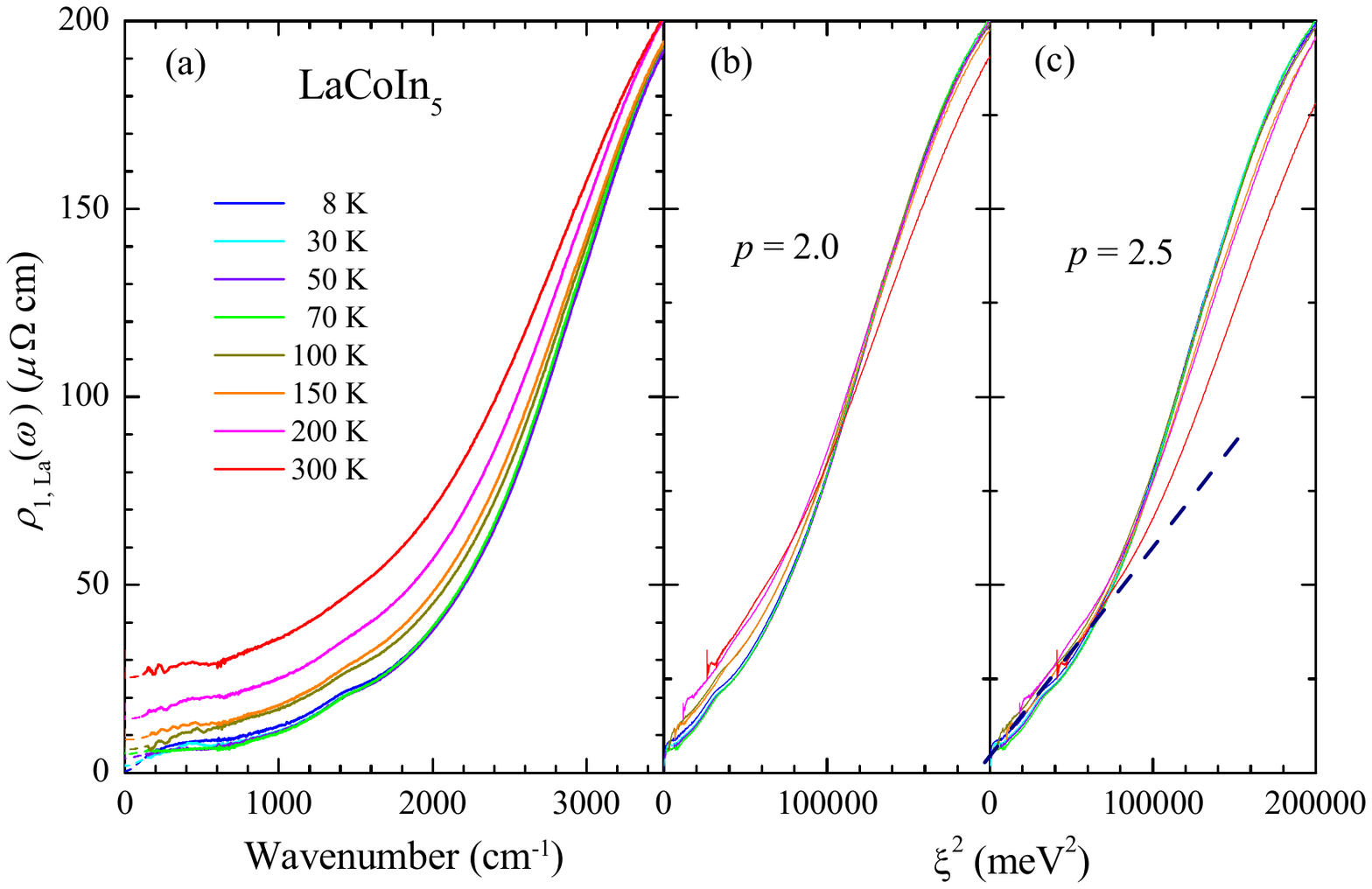}}%
  \vspace*{-0.7 cm}%
\caption{Optical resistivity spectra of (a) LaCoIn$_5$ at various temperatures. (b, c) Optical resistivity spectra of LaCoIn$_5$ as functions of $\xi^2 = (\hbar \omega)^2+(p \pi k_B T)^2$ for the cases of $p =$ 2.0 and 2.5. The dashed line is a guide to the eyes.}
 \label{fig6}
\end{figure}

The optical complex resistivity ($\tilde{\rho}(\omega)$)\cite{nagel:2012} can be defined by the inverse of the complex optical conductivity $\tilde{\sigma}(\omega)$, i.e., $\tilde{\rho}(\omega)\equiv 1/\tilde{\sigma}(\omega)$. The real (imaginary) part of the optical resistivity is closely related to the optical scattering rate (optical mass enhancement factor) in the extended Drude formalism \cite{gotze:1972,allen:1977,puchkov:1996,hwang:2004}. The real parts of the optical resistivity of LaCoIn$_5$ are displayed in Fig. \ref{fig6}(a). Figs. \ref{fig6}(b) and (c) show the optical resistivity spectra of LaCoIn$_5$ as functions of $\xi^2 = (\hbar \omega)^2+(p \pi k_B T)^2$, where $p$ is an adjustable parameter, $\hbar$ is the reduced Planck's constant, and $k_B$ is the Boltzmann constant. This plot has been introduced in literature \cite{mirzaei:2013}. If all curves fall into a single curve in the low $\xi$ region the resistivity presents a Fermi liquid behavior \cite{nagel:2012,mirzaei:2013}. When the parameter $p$ is larger than 2.0 both elastic and inelastic scattering contributes to the optical conductivity \cite{maslov:2012}. Since $p$ is larger than 2.0 for the resistivity of LaCoIn$_5$, charge carriers in LaCoIn$_5$ may experience both elastic and inelastic mechanisms even though the elastic mechanism seems to be dominant against the inelastic one. Therefore, LaCoIn$_5$ deviates from a pure metal with $p =$ 2.0 as shown in Figs. \ref{fig6}(b) and (c) \cite{nagel:2012}. It is worth noting that the curve at 300 K shows a large deviation in the high $\xi^2$ values when compared to other temperature curves. We do not clearly know the reason; we suspect that it is likely not a good data curve.

\begin{figure}[!htbp]
  \vspace*{-0.8 cm}%
  \centerline{\includegraphics[width= 6.0 in]{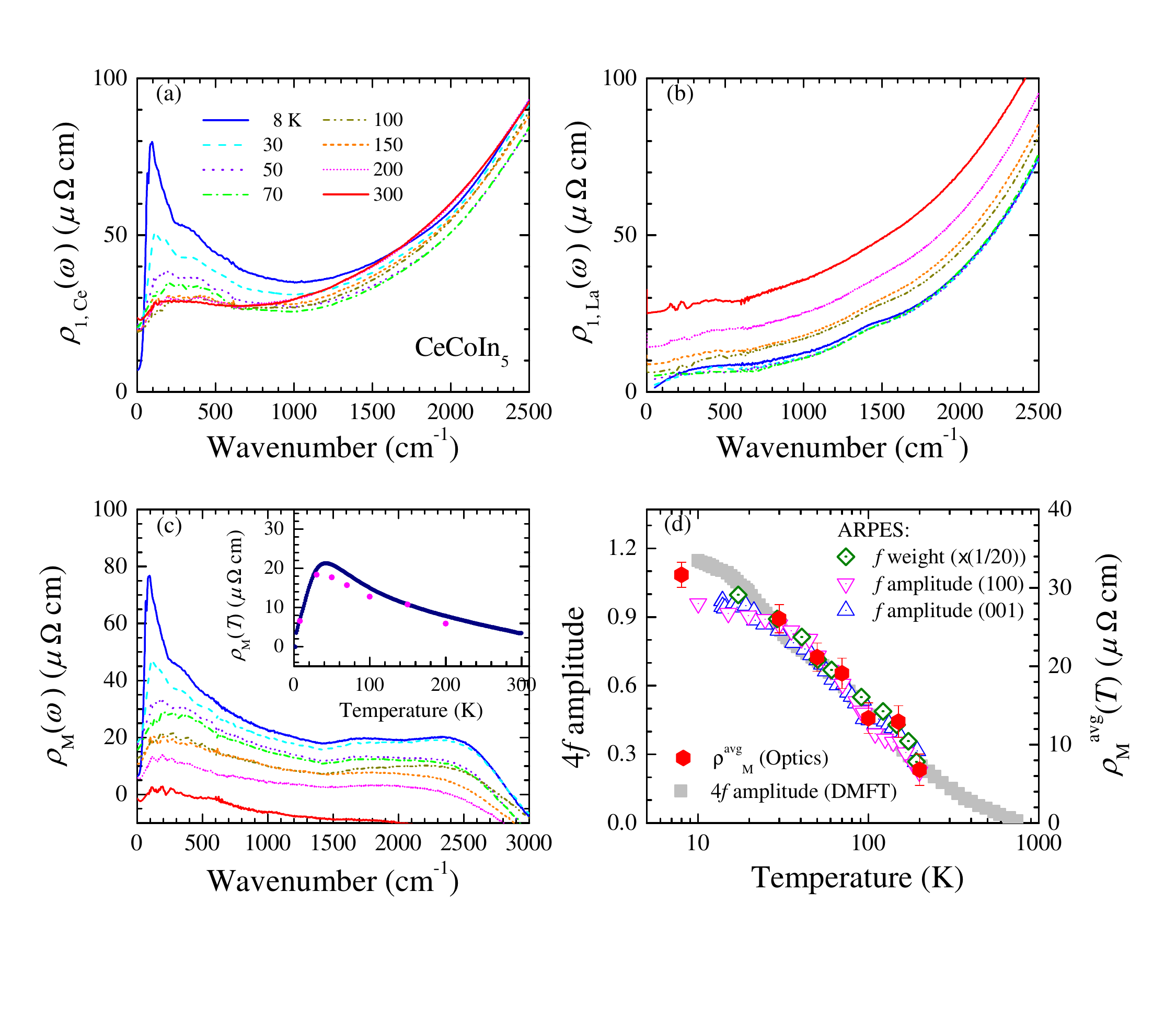}}%
  \vspace*{-0.9 cm}%
\caption{Optical resistivity spectra of (a) CeCoIn$_5$ and (b) LaCoIn$_5$ at various temperatures. (c) The magnetic optical resistivity spectra at various temperatures. (d) The average magnetic resistivity data as a function of temperature. The DMFT calculated 4$f$ density of states (gray solid squares), APRES measured $f$-peak amplitudes in (100) and (001) surfaces \cite{jang:2020}, and the $f$ electron spectral weight from ARPES study \cite{chen:2017}.}
 \label{fig7}
\end{figure}

With a similar idea as the magnetic DC resistivity\cite{nakatsuji:2002}, the magnetic optical resistivity can be defined as $\rho_M(\omega) \equiv \rho_{1,\mathrm{Ce}}(\omega)-\rho_{1,\mathrm{La}}(\omega)$, where $\rho_{1,\mathrm{Ce}}(\omega)$ and $\rho_{1,\mathrm{La}}(\omega)$ are the real parts of the optical resistivity of CeCoIn$_5$ and LaCoIn$_5$, respectively. In Fig. \ref{fig7}(a) and \ref{fig7}(b), we show the optical resistivity spectra of CeCoIn$_5$ and LaCoIn$_5$ at various temperatures. The optical resistivity of CeCoIn$_5$ shows that, as the temperature is lowered from 300 K, a peak near 300 cm$^{-1}$ gradually grows down to $\sim$50 K and then rapidly grows below $\sim$50 K, which is close to the coherence temperature, $T^*$. The peak position shifts to lower energy with decreasing temperature and is located at $\sim$90 cm$^{-1}$ at 8 K. On the other hand, the optical resistivity of LaCoIn$_5$ is a roughly quadratic function of the frequency and temperature, which is a Fermi liquid behavior (see Fig. \ref{fig6} and the related text). We subtract the optical resistivity of LaCoIn$_5$ from that of CeCoIn$_5$ to obtain the magnetic optical resistivity ($\rho_{\mathrm{M}}(\omega,T)$), i.e., $\rho_{\mathrm{M}}(\omega,T) \equiv \rho_{\mathrm{1,Ce}}(\omega,T)-\rho_{\mathrm{1,La}}(\omega,T)$, where $\rho_{\mathrm{1,Ce}}(\omega,T)$ and $\rho_{\mathrm{1,La}}(\omega,T)$ are the optical resistivity spectra of CeCoIn$_5$ and LaCoIn$_5$, respectively (see Fig. S3 in Supplementary Materials). The magnetic optical resistivity can be understood by considering the additional optical resistivity caused by the localized 4$f$ electrons. Fig. \ref{fig7}(c) shows the magnetic optical resistivity spectra ($\rho_{\mathrm{M}}(\omega)$) at various temperatures. The magnetic optical resistivity exhibits strong temperature and frequency dependencies. In the inset of Fig. \ref{fig7}(c), the magnetic DC resistivity ($\rho_{\mathrm{M}}(T)$) is shown. In the inset, we also show the magnetic DC resistivity obtained from extrapolations of the magnetic optical resistivity spectra to zero frequency; the two DC resistivity data sets agree well. Note that the data point at 300 K is below zero, which is not physical. Therefore, we did not include the data point (see Fig. \ref{fig7}(d)) in the following discussion.

In the magnetic DC resistivity, we observe a peak where the Kondo lattice coherence occurs. We also observed a peak in the magnetic optical resistivity, which might be related to the Kondo lattice coherence in the frequency domain. The magnetic DC resistivity is known to be caused by the spins of the localized 4$f$ electrons. Similarly, the magnetic optical resistivity can be attributed to the localized 4$f$ electrons. We take the average of the magnetic optical resistivity up to a frequency ($\Omega_\mathrm{M}$), which gives positive values for the magnetic optical resistivity, i.e., $\rho^{\mathrm{avg}}_{\mathrm{M}}(T) \equiv(1/\Omega_{\mathrm{M}})\int_0^{\Omega_\mathrm{M}}\rho_\mathrm{M}(\omega, T)d\omega$, and display the average magnetic resistivity ($\rho_{\mathrm{M}}^{\mathrm{avg}}$) as a function of temperature in Fig. \ref{fig7}(d). In the figure, we show the normalized amplitude of the 4$f$ density of states (DOS) obtained using the dynamic mean-field theory (DMFT) calculations and the $f$-peak amplitudes in the (100) and (001) planes obtained using ARPES as functions of temperature \cite{jang:2020}. We also show the $f$-electron spectral weight obtained by the ARPES study as a function of temperature \cite{chen:2017}. The resulting average magnetic resistivity exhibits a temperature-dependent trend similar to those of the DMFT calculated 4$f$ density of state, the ARPES $f$-peak amplitudes \cite{jang:2020}, and the ARPES $f$-electron spectral weight \cite{chen:2017}. All the ARPES and optics data, including the DMFT result, roughly fall into a single curve, indicating that the magnetic optical resistivity is closely related to the 4$f$ electrons in CeCoIn$_5$. We note that Fig. \ref{fig7}(d) on the linear scale of the horizontal axis is displayed in Fig. S4 of Supplementary Materials to show the data more clearly. The previous ARPES and our optical results, along with the DMFT calculations, suggest that the 4$f$ electron effects may exist at temperatures above the known onset temperature of Kondo scattering, $T_K'$. Another suggestive way to obtain the magnetic optical resistivity can be done by fitting the optical resistivity of CeCoIn$_5$ in a high energy region with a quadratic function of frequency, i.e., $\rho^{\mathrm{FL}}(\omega, T) = A(T)\omega^2+B(T)$, where $\rho^{\mathrm{FL}}(\omega, T)$ is the Fermi liquid optical resistivity, and $A(T)$ and $B(T)$ are adjustable parameters (see Fig. S5 in Supplementary Materials). It is worth noting that because the LaCoIn$_5$ is a Fermi liquid, we assume that the background is in a Fermi liquid phase. We obtain the magnetic optical resistivity spectra by subtracting the Fermi liquid background from the measured optical resistivity of CeCoIn$^{5}$ and the resulting magnetic optical resistivity spectra are shown in Fig. S6(a) in Supplementary Materials. In Fig. S6(b), we also show the average magnetic resistivity as a function of the temperature, which is a similar temperature dependence as that in Fig. \ref{fig7}(d).

\subsection*{Extended Drude model and optical effective mass}

We applied the extended Drude model formalism to obtain the information on the electron-electron correlations. The optical conductivity can be expressed in the extended Drude model formalism as \cite{gotze:1972,allen:1977,puchkov:1996,hwang:2004}
\begin{eqnarray}
  \tilde{\sigma}(\omega)&=&i\frac{\Omega_p^2}{4\pi}\frac{1}{[m^*_{op}(\omega)/m_b]\omega+i[1/\tau^{op}(\omega)]} \\ \nonumber
  &=&i\frac{\Omega_p^2}{4\pi} \frac{1}{\omega+[-2\tilde{\Sigma}^{op}(\omega)]}
\end{eqnarray}
where $\Omega_p$ is the plasma frequency of the charge carriers, $1/\tau^{op}(\omega)$ (= $\Im(i\Omega_p^2/[4\pi \tilde{\sigma}(\omega)]$)) is the optical (or inelastic) scattering rate, and $m^*_{op}(\omega)/m_b$ (= $\Re(i\Omega_p^2/[4\pi \omega \:\tilde{\sigma}(\omega)]$)) is the optical effective mass with respect to the band mass ($m_b$) caused by the correlations between electrons. To include all charge carriers, we used an integrated spectral weight up to 2500 cm$^{-1}$ as in literature\cite{singley:2002} and estimated $\Omega_p$. The estimated plasma frequency is 29700 cm$^{-1}$, which is corresponding to 1.15 carriers per formula unit. $\tilde{\Sigma}^{op}(\omega)$ is the complex optical self-energy, which is an optical quantity that corresponds to the quasiparticle self-energy\cite{hwang:2004}, and also contains information about the correlations; $-2\Sigma^{op}_1(\omega)=[m^*_{op}(\omega)/m_b-1]\omega$ and $-2\Sigma^{op}_2(\omega)=1/\tau^{op}(\omega)$. The real and imaginary parts of the optical self-energy form a Kramers-Kronig pair \cite{hwang:2004,hwang:2020}. In particular, the real part of the self-energy is directly related to the band renormalization caused by the correlations. We note that the ideal extended Drude model formalism is for a single band with an infinite bandwidth \cite{wu:2010}, but the correlation effects on charge carriers are dominant in the low-frequency region near the Fermi level, that is, $\omega = 0$. Therefore, the extended Drude model can be applied to an averaged band of a material system with multiple bands at the Fermi level in the low-frequency region near $\omega =$ 0 \cite{singley:2002}.

\begin{figure}[!htbp]
  \vspace*{-0.9 cm}%
  \centerline{\includegraphics[width= 6.0 in]{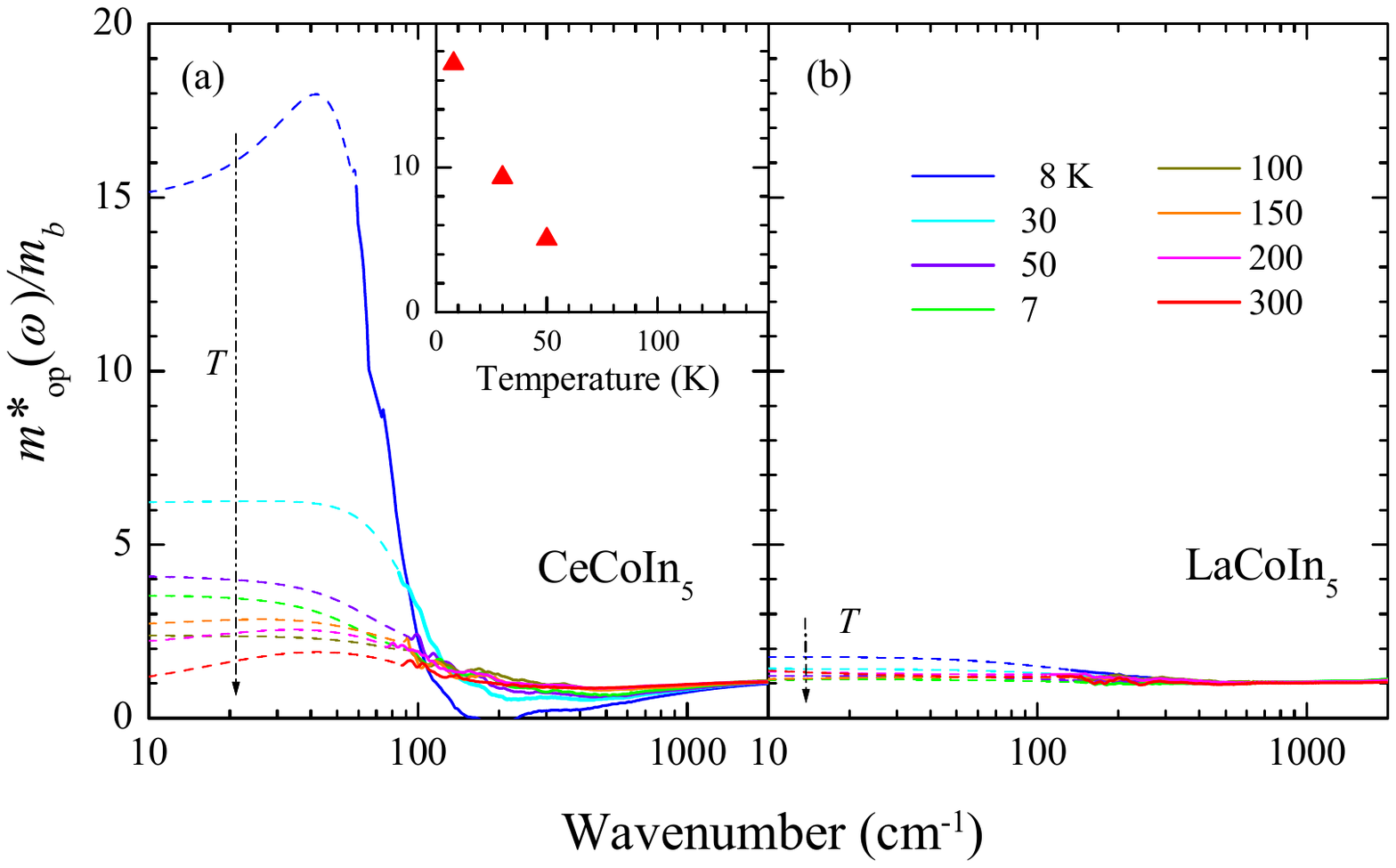}}%
  \vspace*{-1.1 cm}%
\caption{Optical effective masses with respect to the band mass of (a) CeCoIn$_5$ and (b) LaCoIn$_5$ at various temperatures in a frequency range up to 2500 cm$^{-1}$. The inset shows the effective masses with respect to the band mass as a function of temperature (see in the text for a detailed description).}
 \label{fig8}
\end{figure}

Because the optical scattering rate ($1/\tau^{op}(\omega)$) is proportional to the real part of the optical resistivity ($\rho_1(\omega)$), we concentrate on the optical effective mass with respect to the band mass. Figs. \ref{fig8}(a) and (b) show the optical effective masses of CeCoIn$_5$ and LaCoIn$_5$ at various temperatures, respectively. CeCoIn$_5$ shows a dramatic temperature dependence in the low-frequency region near zero. We estimated the effective mass at zero frequency by an extrapolation. The estimated effective mass is approximately 1.0 at 300 K. It increases upon cooling down the sample and becomes $\sim$15 at 8 K. The effective mass at zero energy can also be obtained from the spectral weight redistribution caused by the correlations. In general, the spectral weight is divided into two components (coherence and incoherence) by the correlations. The division can appear as a dip in the optical conductivity if the correlations originate from coupling to a sharp bosonic mode\cite{hwang:2008a}. A previous study \cite{singley:2002} has introduced a method that can be used to estimate the effective mass using the spectral weight redistribution by the correlations, i.e., $m^{*}(T)/m_b = \int_0^{\omega_c^{300 K}}\sigma_1(\omega, 300 K)d\omega/\int_0^{\omega_c^{T}}\sigma_1(\omega, T)d\omega$, where $\omega_c^{300 K}$ and $\omega_c^{T}$ are the cutoff frequencies at 300 K and $T$, respectively. We took 2500 cm$^{-1}$and the frequency at the dip ($\sim$ 30 cm$^{-1}$) below the hybridization gap (see Fig. 2(a)) as the cutoff frequencies at 300 K and $T$, respectively. In the inset of Fig. \ref{fig8}(a), we show the effective masses of CeCoIn$_5$ estimated by using the spectral weight redistribution. The two results obtained using the two different methods more or less agreed. We note that, at high temperatures, we cannot clearly determine the energy (or the dip) that divides the coherent and incoherent components of the spectral weight. Therefore, we could not obtain the effective masses above 50 K using spectral redistribution. On the other hand, the optical effective masses of LaCoIn$_5$ are lower than 2.0 in the low-frequency region and exhibit small frequency and temperature dependencies. We note that the large effective mass originated from the band renormalization caused by the hybridization between the conduction electrons (In 5$p$ electrons) and the local moments (Ce 4$f$ electrons). The temperature dependence of the effective mass comes from the temperature dependence of the strength of the hybridization.

\section*{Conclusions}

We performed a comparative study of CeCoIn$_5$ with LaCoIn$_5$ using an infrared spectroscopic technique. These two material systems have the same crystal structure but different electron configurations. CeCoIn$_5$ has one more electron in the 4$f$ orbital. We found that these two material systems exhibited significantly different temperature-dependent spectroscopic properties near the Fermi surface. To present the difference, we introduced and used the complex optical resistivity and magnetic optical resistivity. The magnetic optical resistivity exhibited additional resistivity in CeCoIn$_5$ compared to that of LaCoIn$_5$. Therefore, the magnetic optical resistivity could be interpreted as the contribution of the 4$f$ electrons in CeCoIn$_5$ and exhibited strong temperature dependence, which agree with the temperature-dependent density of states of 4$f$ electrons obtained by DMFT calculations and temperature-dependent $f$-amplitudes of measured ARPES data \cite{jang:2020, chen:2017}. We found that the onset temperature of the Kondo effect is higher than the known onset temperature of Kondo scattering ($T_K'$), which is consistent with the recent combined study of DMFT and ARPES \cite{jang:2020} and the previous ARPES study \cite{chen:2017}. We also obtained the hybridization gap distribution function of CeCoIn$_5$ using a proposed approach \cite{burch:2007} through a maximum entropy method. We found that the distribution function mainly consisted of two (small and large) components. We assigned the small and large gaps to the in-plane and out-of-plane hybridization gaps, respectively. We also used the extended Drude model formalism to extract information on strong electron-electron correlations. As we expected, CeCoIn$_5$ showed much larger effective masses compared with LaCoIn$_5$. Our results demonstrate that the $f$ electron system exhibits significantly different electronic properties compared with the system without $f$ electrons near the Fermi level at low temperatures. The $f$ electron effect gradually builds up with lowering the temperature, and the onset temperature of the $f$ electron effect is high above $T_K'$. We expect that our finding will help us understand the electronic structure evolution of CeCoIn$_5$ with temperature. \\ \\

\noindent {\bf Acknowledgements} This paper was supported by the National Research Foundation of Korea (NRFK Grant No. 2017R1A2B4007387, 2021R1A2C2010925, and 2021R1A2C101109811). This research was also supported by the BrainLink program, funded by the Ministry of Science and ICT through the National Research Foundation of Korea (2022H1D3A3A01077468).
\\
%
%
\bibliographystyle{naturemag}
\bibliography{bib}

\begin{thebibliography}{10}
\expandafter\ifx\csname url\endcsname\relax
  \def\url#1{\texttt{#1}}\fi
\expandafter\ifx\csname urlprefix\endcsname\relax\def\urlprefix{URL }\fi
\providecommand{\bibinfo}[2]{#2}
\providecommand{\eprint}[2][]{\url{#2}}

\bibitem{steglich:1979}
\bibinfo{author}{Steglich, F.} \emph{et~al.}
\newblock \bibinfo{title}{Superconductivity in the presence of strong pauli
  paramagnetism: \mbox{CeCu$_2$Si$_2$}}.
\newblock \emph{\bibinfo{journal}{Phys. Rev. Lett.}}
  \textbf{\bibinfo{volume}{43}}, \bibinfo{pages}{1892} (\bibinfo{year}{1979}).

\bibitem{bonn:1988}
\bibinfo{author}{Bonn, D.~A.}, \bibinfo{author}{Garrett, J.~D.} \&
  \bibinfo{author}{Timusk, T.}
\newblock \bibinfo{title}{Far-infrared properties of \mbox{URu$_2$Si$_2$}}.
\newblock \emph{\bibinfo{journal}{Phys. Rev. Lett.}}
  \textbf{\bibinfo{volume}{61}}, \bibinfo{pages}{1305} (\bibinfo{year}{1988}).

\bibitem{donovan:1997}
\bibinfo{author}{Donovan, S.}, \bibinfo{author}{Schwartz, A.} \&
  \bibinfo{author}{Gruner, G.}
\newblock \bibinfo{title}{Observation of an optical pseudogap in
  \mbox{UPt$_3$}}.
\newblock \emph{\bibinfo{journal}{Phys. Rev. Lett.}}
  \textbf{\bibinfo{volume}{79}}, \bibinfo{pages}{1401} (\bibinfo{year}{1997}).

\bibitem{degiorgi:1999}
\bibinfo{author}{Degiorgi, L.}
\newblock \bibinfo{title}{The electrodynamic response of heavy-electron
  compounds}.
\newblock \emph{\bibinfo{journal}{Rev. Mod. Phys.}}
  \textbf{\bibinfo{volume}{71}}, \bibinfo{pages}{687} (\bibinfo{year}{1999}).

\bibitem{petrovic:2001}
\bibinfo{author}{Petrovic, C.} \emph{et~al.}
\newblock \bibinfo{title}{Heavy-fermion superconductivity in \mbox{CeCoIn$_5$
  at 2.3 K}}.
\newblock \emph{\bibinfo{journal}{J. Phys.: Condens. Matter}}
  \textbf{\bibinfo{volume}{13}}, \bibinfo{pages}{L337} (\bibinfo{year}{2001}).

\bibitem{dordevic:2001}
\bibinfo{author}{Dordevic, S.~V.}, \bibinfo{author}{Basov, D.~N.},
  \bibinfo{author}{Dilley, N.~R.}, \bibinfo{author}{Bauer, E.~D.} \&
  \bibinfo{author}{Maple, M.~B.}
\newblock \bibinfo{title}{Hybridization gap in heavy fermion compounds}.
\newblock \emph{\bibinfo{journal}{Phys. Rev. Lett.}}
  \textbf{\bibinfo{volume}{86}}, \bibinfo{pages}{684} (\bibinfo{year}{2001}).

\bibitem{dressel:2002}
\bibinfo{author}{Dressel, M.} \emph{et~al.}
\newblock \bibinfo{title}{Nature of heavy quasiparticles in magnetically
  ordered heavy fermions \mbox{UPd$_2$Al$_3$} and \mbox{UPt$_3$}}.
\newblock \emph{\bibinfo{journal}{Phys. Rev. Lett.}}
  \textbf{\bibinfo{volume}{88}}, \bibinfo{pages}{186404}
  (\bibinfo{year}{2002}).

\bibitem{hancock:2004}
\bibinfo{author}{Hancock, J.~N.}, \bibinfo{author}{McKnew, T.},
  \bibinfo{author}{Schlesinger, Z.}, \bibinfo{author}{Sarrao, J.~L.} \&
  \bibinfo{author}{Fisk, Z.}
\newblock \bibinfo{title}{Kondo scaling in the optical response of
  \mbox{YbIn$_{1-x}$Ag$_x$Cu$_4$}}.
\newblock \emph{\bibinfo{journal}{Phys. Rev. Lett.}}
  \textbf{\bibinfo{volume}{92}}, \bibinfo{pages}{186405}
  (\bibinfo{year}{2004}).

\bibitem{silhanek:2006}
\bibinfo{author}{Silhanek, A.~V.} \emph{et~al.}
\newblock \bibinfo{title}{Nonlocal magnetic field-tuned quantum criticality in
  cubic \mbox{CeIn$_{3-x}$Sn$_x$ ($x =$ 0.25)}}.
\newblock \emph{\bibinfo{journal}{Phys. Rev. Lett.}}
  \textbf{\bibinfo{volume}{96}}, \bibinfo{pages}{206401}
  (\bibinfo{year}{2006}).

\bibitem{park:2008}
\bibinfo{author}{Park, T.} \emph{et~al.}
\newblock \bibinfo{title}{Isotropic quantum scattering and unconventional
  superconductivity}.
\newblock \emph{\bibinfo{journal}{Nature}} \textbf{\bibinfo{volume}{456}},
  \bibinfo{pages}{366} (\bibinfo{year}{2008}).

\bibitem{nagel:2012}
\bibinfo{author}{Nagel, U.} \emph{et~al.}
\newblock \bibinfo{title}{Optical spectroscopy shows that the normal state of
  \mbox{URu$_2$Si$_2$} is an anomalous fermi liquid}.
\newblock \emph{\bibinfo{journal}{PNAS}} \textbf{\bibinfo{volume}{109}},
  \bibinfo{pages}{19161} (\bibinfo{year}{2012}).

\bibitem{wirth:2016}
\bibinfo{author}{Wirth, S.} \& \bibinfo{author}{Steglich, F.}
\newblock \bibinfo{title}{Exploring heavy fermions from macroscopic to
  microscopic length scales}.
\newblock \emph{\bibinfo{journal}{Nat. Rev.}} \textbf{\bibinfo{volume}{1}},
  \bibinfo{pages}{1} (\bibinfo{year}{2016}).

\bibitem{chen:2016}
\bibinfo{author}{Chen, R.~Y.} \& \bibinfo{author}{Wang, N.~L.}
\newblock \bibinfo{title}{Kondo scaling in the optical response of
  \mbox{YbIn$_{1-x}$Ag$_x$Cu$_4$}}.
\newblock \emph{\bibinfo{journal}{Reports on Progress in Physics}}
  \textbf{\bibinfo{volume}{79}}, \bibinfo{pages}{064502}
  (\bibinfo{year}{2016}).

\bibitem{kirchner:2020}
\bibinfo{author}{Kirchner, S.} \emph{et~al.}
\newblock \bibinfo{title}{Colloquium: Heavy-electron quantum criticality and
  single-particle spectroscopy}.
\newblock \emph{\bibinfo{journal}{Rev. Mod. Phys.}}
  \textbf{\bibinfo{volume}{92}}, \bibinfo{pages}{011002}
  (\bibinfo{year}{2020}).

\bibitem{doniach:1977}
\bibinfo{author}{Doniach, S.}
\newblock \bibinfo{title}{The kondo lattice and weak antiferromagnetism}.
\newblock \emph{\bibinfo{journal}{Physica B}} \textbf{\bibinfo{volume}{91}},
  \bibinfo{pages}{231} (\bibinfo{year}{1977}).

\bibitem{jang:2020}
\bibinfo{author}{Jang, S.} \emph{et~al.}
\newblock \bibinfo{title}{Evolution of the kondo lattice electronic structure
  above the transport coherence temperature}.
\newblock \emph{\bibinfo{journal}{PNAS}} \textbf{\bibinfo{volume}{117}},
  \bibinfo{pages}{23467} (\bibinfo{year}{2020}).

\bibitem{yang:2008}
\bibinfo{author}{Yang, Y.-F.}, \bibinfo{author}{Fisk, Z.},
  \bibinfo{author}{Lee, H.-O.}, \bibinfo{author}{Thompson, J.~D.} \&
  \bibinfo{author}{Pines, D.}
\newblock \bibinfo{title}{Scaling the kondo lattice}.
\newblock \emph{\bibinfo{journal}{Nature}} \textbf{\bibinfo{volume}{454}},
  \bibinfo{pages}{611} (\bibinfo{year}{2008}).

\bibitem{knebel:2001}
\bibinfo{author}{Knebel, G.}, \bibinfo{author}{Braithwaite, D.},
  \bibinfo{author}{Canfield, P.~C.}, \bibinfo{author}{Lapertot, G.} \&
  \bibinfo{author}{Flouquet, J.}
\newblock \bibinfo{title}{Electronic properties of \mbox{CeIn$_3$} under high
  pressure near the quantum critical point}.
\newblock \emph{\bibinfo{journal}{Phys. Rev. B}} \textbf{\bibinfo{volume}{65}},
  \bibinfo{pages}{024425} (\bibinfo{year}{2001}).

\bibitem{sidorov:2002}
\bibinfo{author}{Sidorov, V.~A.} \emph{et~al.}
\newblock \bibinfo{title}{Superconductivity and quantum criticality in
  \mbox{CeCoIn$_5$}}.
\newblock \emph{\bibinfo{journal}{Phys. Rev. Lett.}}
  \textbf{\bibinfo{volume}{89}}, \bibinfo{pages}{157004}
  (\bibinfo{year}{2002}).

\bibitem{singley:2002}
\bibinfo{author}{Singley, E.~J.}, \bibinfo{author}{Basov, D.~N.},
  \bibinfo{author}{Bauer, E.~D.} \& \bibinfo{author}{Maple, M.~B.}
\newblock \bibinfo{title}{Optical conductivity of the heavy fermion
  superconductor \mbox{CeCoIn$_5$}}.
\newblock \emph{\bibinfo{journal}{Phys. Rev. B}} \textbf{\bibinfo{volume}{65}},
  \bibinfo{pages}{161101} (\bibinfo{year}{2002}).

\bibitem{shishido:2002}
\bibinfo{author}{Shishido, H.} \emph{et~al.}
\newblock \bibinfo{title}{Fermi surface, magnetic and superconducting
  properties of \mbox{LaRhIn$_5$} and \mbox{CeTIn$_5$ (T: Co, Rh and Ir)}}.
\newblock \emph{\bibinfo{journal}{J. Phys. Soc. Jpn}}
  \textbf{\bibinfo{volume}{71}}, \bibinfo{pages}{162} (\bibinfo{year}{2002}).

\bibitem{mena:2005}
\bibinfo{author}{Mena, F.~P.}, \bibinfo{author}{van~der Marel, D.} \&
  \bibinfo{author}{Sarrao, J.~L.}
\newblock \bibinfo{title}{Optical conductivity of \mbox{CeMIn$_5$ (M=Co, Rh,
  Ir)}}.
\newblock \emph{\bibinfo{journal}{Phys. Rev. B}} \textbf{\bibinfo{volume}{72}},
  \bibinfo{pages}{045119} (\bibinfo{year}{2005}).

\bibitem{shim:2007}
\bibinfo{author}{Shim, J.~H.}, \bibinfo{author}{Haule, K.} \&
  \bibinfo{author}{Kotliar, G.}
\newblock \bibinfo{title}{Modeling the localized-to-itinerant electronic
  transition in the heavy fermion system \mbox{CeIrIn$_5$}}.
\newblock \emph{\bibinfo{journal}{Science}} \textbf{\bibinfo{volume}{318}},
  \bibinfo{pages}{1615} (\bibinfo{year}{2007}).

\bibitem{burch:2007}
\bibinfo{author}{Burch, K.~S.} \emph{et~al.}
\newblock \bibinfo{title}{Optical signatures of momentum-dependent
  hybridization of the local moments and conduction electrons in kondo
  lattices}.
\newblock \emph{\bibinfo{journal}{Phys. Rev. B}} \textbf{\bibinfo{volume}{75}},
  \bibinfo{pages}{054523} (\bibinfo{year}{2007}).

\bibitem{nakatsuji:2003}
\bibinfo{author}{Nakatsuji, S.}, \bibinfo{author}{Pines, D.} \&
  \bibinfo{author}{Fisk, Z.}
\newblock \bibinfo{title}{Probing the kondo lattice}.
\newblock \emph{\bibinfo{journal}{arXiv preprint arXiv:0304.587v1}}
  (\bibinfo{year}{2003}).

\bibitem{chen:2017}
\bibinfo{author}{Chen, Q.~Y.} \emph{et~al.}
\newblock \bibinfo{title}{Direct observation of how the heavy-fermion state
  develops in \mbox{CeCoIn$_5$}}.
\newblock \emph{\bibinfo{journal}{Phys. Rev. B}} \textbf{\bibinfo{volume}{96}},
  \bibinfo{pages}{045107} (\bibinfo{year}{2017}).

\bibitem{chen:2019}
\bibinfo{author}{Chen, Q.~Y.} \emph{et~al.}
\newblock \bibinfo{title}{Electronic structure study of \mbox{LaCoIn$_5$} and
  its comparison with \mbox{CeCoIn$_5$}}.
\newblock \emph{\bibinfo{journal}{Phys. Rev. B}}
  \textbf{\bibinfo{volume}{100}}, \bibinfo{pages}{035117}
  (\bibinfo{year}{2019}).

\bibitem{gotze:1972}
\bibinfo{author}{G\"{o}tze, W.} \& \bibinfo{author}{W\"{o}lfle, P.}
\newblock \bibinfo{title}{Homogeneous dynamical conductivity of simple metals}.
\newblock \emph{\bibinfo{journal}{Phys. Rev. B}} \textbf{\bibinfo{volume}{6}},
  \bibinfo{pages}{1226} (\bibinfo{year}{1972}).

\bibitem{allen:1977}
\bibinfo{author}{Allen, J.~W.} \& \bibinfo{author}{Mikkelsen, J.~C.}
\newblock \bibinfo{title}{Optical properties of \mbox{CrSb}, \mbox{MnSb},
  \mbox{NiSb}, and \mbox{NiAs}}.
\newblock \emph{\bibinfo{journal}{Phys. Rev. B}} \textbf{\bibinfo{volume}{15}},
  \bibinfo{pages}{2952} (\bibinfo{year}{1977}).

\bibitem{puchkov:1996}
\bibinfo{author}{Puchkov, A.~V.}, \bibinfo{author}{Basov, D.~N.} \&
  \bibinfo{author}{Timusk, T.}
\newblock \bibinfo{title}{The pseudogap state in high-\mbox{T$_c$}
  superconductors: an infrared study}.
\newblock \emph{\bibinfo{journal}{J. Phys.: Cond. Matter}}
  \textbf{\bibinfo{volume}{8}}, \bibinfo{pages}{10049} (\bibinfo{year}{1996}).

\bibitem{hwang:2004}
\bibinfo{author}{Hwang, J.}, \bibinfo{author}{Timusk, T.} \&
  \bibinfo{author}{Gu, G.~D.}
\newblock \bibinfo{title}{High-transition-temperature superconductivity in the
  absence of the magnetic-resonance mode}.
\newblock \emph{\bibinfo{journal}{Nature (London)}}
  \textbf{\bibinfo{volume}{427}}, \bibinfo{pages}{714} (\bibinfo{year}{2004}).

\bibitem{hu:2013}
\bibinfo{author}{Hu, T.} \emph{et~al.}
\newblock \bibinfo{title}{Non-fermi liquid regimes with and without quantum
  criticality in \mbox{Ce$_{1-x}$Yb$_x$CoIn$_5$}}.
\newblock \emph{\bibinfo{journal}{PNAS}} \textbf{\bibinfo{volume}{110}},
  \bibinfo{pages}{7160} (\bibinfo{year}{2013}).

\bibitem{macaluso:2002}
\bibinfo{author}{Macaluso, R.~T.} \emph{et~al.}
\newblock \bibinfo{title}{Crystal growth and structure determination of
  \mbox{La$M$In$_5$ ($M$ = Co, Rh, Ir)}}.
\newblock \emph{\bibinfo{journal}{J. Sol. State. Chem.}}
  \textbf{\bibinfo{volume}{166}}, \bibinfo{pages}{245} (\bibinfo{year}{2002}).

\bibitem{homes:1993}
\bibinfo{author}{Homes, C.~C.}, \bibinfo{author}{Reedyk, M.~A.},
  \bibinfo{author}{Crandles, D.~A.} \& \bibinfo{author}{Timusk, T.}
\newblock \bibinfo{title}{Technique for measuring the reflectance of irregular,
  submillimeter-sized samples}.
\newblock \emph{\bibinfo{journal}{Appl. Opt.}} \textbf{\bibinfo{volume}{32}},
  \bibinfo{pages}{2976} (\bibinfo{year}{1993}).

\bibitem{nakatsuji:2002}
\bibinfo{author}{Nakatsuji, S.} \emph{et~al.}
\newblock \bibinfo{title}{Intersite coupling effects in a kondo lattice}.
\newblock \emph{\bibinfo{journal}{Phys. Rev. Lett.}}
  \textbf{\bibinfo{volume}{89}}, \bibinfo{pages}{106402}
  (\bibinfo{year}{2002}).

\bibitem{singley:2001}
\bibinfo{author}{Singley, E.~J.}, \bibinfo{author}{Basov, D.~N.},
  \bibinfo{author}{Kurahashi, K.}, \bibinfo{author}{Uefuji, T.} \&
  \bibinfo{author}{Yamada, K.}
\newblock \emph{\bibinfo{journal}{Phys. Rev. B}} \textbf{\bibinfo{volume}{64}},
  \bibinfo{pages}{224503} (\bibinfo{year}{2001}).

\bibitem{allen:1971}
\bibinfo{author}{Allen, P.~B.}
\newblock \bibinfo{title}{Electron-phonon effects in the infrared properties of
  metals}.
\newblock \emph{\bibinfo{journal}{Phys. Rev. B}} \textbf{\bibinfo{volume}{3}},
  \bibinfo{pages}{305} (\bibinfo{year}{1971}).

\bibitem{schachinger:2006}
\bibinfo{author}{Schachinger, E.}, \bibinfo{author}{Neuber, D.} \&
  \bibinfo{author}{Carbotte, J.~P.}
\newblock \bibinfo{title}{Inversion techniques for optical conductivity data}.
\newblock \emph{\bibinfo{journal}{Phys. Rev. B}} \textbf{\bibinfo{volume}{73}},
  \bibinfo{pages}{184507} (\bibinfo{year}{2006}).

\bibitem{mirzaei:2013}
\bibinfo{author}{Mirzaei, S.~I.} \emph{et~al.}
\newblock \bibinfo{title}{Spectroscopic evidence for fermi liquid-like energy
  and temperature dependence of the relaxation rate in the pseudogap phase of
  the cuprates}.
\newblock \emph{\bibinfo{journal}{PNAS}} \textbf{\bibinfo{volume}{110}},
  \bibinfo{pages}{5774} (\bibinfo{year}{2013}).

\bibitem{maslov:2012}
\bibinfo{author}{Maslov, D.~L.} \& \bibinfo{author}{Chubukov, A.~V.}
\newblock \bibinfo{title}{First-matsubara-frequency rule in a fermi liquid. ii.
  optical conductivity and comparison to experiment}.
\newblock \emph{\bibinfo{journal}{Phys. Rev. B}} \textbf{\bibinfo{volume}{86}},
  \bibinfo{pages}{155137} (\bibinfo{year}{2012}).

\bibitem{hwang:2020}
\bibinfo{author}{Hwang, J.}
\newblock \bibinfo{title}{Extended drude model analysis of the optical spectra
  of correlated electron systems in a d-wave superconducting state}.
\newblock \emph{\bibinfo{journal}{Journal of Korean Physical Society}}
  \textbf{\bibinfo{volume}{76}}, \bibinfo{pages}{736} (\bibinfo{year}{2020}).

\bibitem{wu:2010}
\bibinfo{author}{Wu, D.} \emph{et~al.}
\newblock \bibinfo{title}{Eliashberg analysis of optical spectra reveals a
  strong coupling of charge carriers to spin fluctuations in doped
  iron-pnictide \mbox{BaFe$_2$As$_2$}superconductors}.
\newblock \emph{\bibinfo{journal}{Phys. Rev. B}} \textbf{\bibinfo{volume}{82}},
  \bibinfo{pages}{144519} (\bibinfo{year}{2010}).

\bibitem{hwang:2008a}
\bibinfo{author}{Hwang, J.}, \bibinfo{author}{Yang, J.},
  \bibinfo{author}{Carbotte, J.~P.} \& \bibinfo{author}{Timusk, T.}
\newblock \bibinfo{title}{Manifestation of the pseudogap in ab-plane optical
  characteristics}.
\newblock \emph{\bibinfo{journal}{J. Phys. Condens. Matter}}
  \textbf{\bibinfo{volume}{20}}, \bibinfo{pages}{295215}
  (\bibinfo{year}{2008}).

\end{thebibliography}

\end{document}